\documentclass[12pt]{article}
\usepackage{amsmath,amsthm,amsfonts,amssymb,amscd}
\usepackage{graphics}
\usepackage[latin1]{inputenc}

\renewcommand{\thefootnote}

\begin{document}

\centerline{\large\bf A gauge invariant path integral}

\smallskip

\centerline{\large\bf for Electrodynamics with Magnetic Monopoles}

\smallskip

\centerline{\large\bf in the Hestenes-Haddamard-Rodrigues formalism}

\vglue .4in

\centerline{\large\bf Luiz C.L. Botelho}

\vglue .3in

\centerline{Departamento de Matemática Aplicada}

\smallskip

\centerline{Instituto de Matemática, Universidade Federal Fluminense}

\smallskip

\centerline{Rua Mario Santos Braga, CEP 24220-140}

\smallskip

\centerline{Niterói, Rio de Janeiro, Brasil}

\smallskip

\centerline{e-mail: botelho.luiz@superig.com.br}

\vglue .5in

\noindent\textbf{Abstract:} We propose a new path integral for QED in presence of magnetic monopoles on the formalism of Geometric Algebra of Hestenes-Haddamard-Rodrigues written in terms of Dirac matrixes.

\footnote{Presently a CNPq visiting research at IMEEC-UNICAMP.}

\vglue .2in

\noindent{\large\bf 1.\, An Euclidean Path Integral for Magnetic Monopoles}

\vglue .3in

One of the most appealing question on quantum field theory is an old  question posed by P.A.M. Dirac: ``Do we have a quantum field theory of magnetic monopoles?''.

\bigskip

Unfortunately the general answer to the question is no. What has been studied in the literature is the introduction of magnetic monopoles as external backgrounds or strings, due to its supposed massiveness (when in velocity the monopole is equivalent to a Rinocherus in full  speed!), which turns out its nature to be a semi-classical level phenomena ([1]).

In this short comment directed to an Theoretical Physicist's audience, we wish to point out that it is possible to second quantize the monopole electromagnetic field in a gauge invariant way through the use of the techniques of Geometric Algebra [2], but now written in simple terms of the physicists elementary Dirac Algebra.

Let us start our note by recalling the mathematical method of introducing magnetic monopoles without unphysical strings ([1]) in the Haddamard-Hestenes-Rodrigues approach, but now re-witten in terms of elementary Dirac matrixes  ([2], [3]). In the tensor calculus notation, the Maxwell equations on $R^4$ in the presence of an electric source $j^\nu(x)$ and magnetic monopole source $k^\nu(x)$ (both supposed divergenceless) write as of a
\begin{align}
&(\partial_\mu\,F^{\mu\nu})(x) = j^\nu(x) \tag{1}\\
&\big(\partial_\mu(^*F^{\mu\nu})\big)(x) = k^\nu(x) \tag{2}
\end{align}
plus the usual Sommerfeld radiation conditions at the $\mathbb{R}^4$ infinity imposed now for the magnetic monopole electric charge generated Electromagnetic strenght fields.

This classical problem is addressed in the Geometric algebra formalism by the introduction of a complex matrix $[F](x)$ on the Dirac Algebra of matrixes, and an antisymmetric rank two tensor with the following relationship with the electromagnetic strenght fields
\begin{equation}
F_{\mu\nu}(x) = \varepsilon^{\mu\nu\rho\sigma}\, B_{\sigma\rho}+ \frac 14\, {\rm Tr}(\gamma^\mu[F]\gamma^\nu) \tag{3}
\end{equation}
Here ours proposed tensorial potential dynamical equation reads as of as
\begin{equation}
\big(\varepsilon^{\mu\nu\rho\sigma}\, \partial_\mu\,B_{\sigma\rho}\big)(x) = 0 \tag{4}
\end{equation}
\begin{equation}
\big(\partial_\mu\,B^{\mu\nu}\big)(x) = k^\nu(x) - \frac 18\, \overbrace{\varepsilon^{\mu\nu\sigma\rho} \partial_\mu\big\{{\rm Tr}_{\rm Dirac} (\gamma^\sigma |[F]|^\rho)\big\}}^{\overline{j}^\nu \equiv}\tag{5}
\end{equation}

Note that the set of equations eq(4)-eq(5) can be elementarly solved as linear functional of a monopole divergenceless current sources $k^\gamma$ and $\overline{j}^\gamma$.
\begin{equation}
B_{\sigma\rho}(x) = (\partial_\sigma Q_\rho - \partial_\rho Q_\sigma)(x). 
\tag{6}
\end{equation}
It reads as of as in full.
\begin{equation}
B_{\sigma\rho}(x) = (\partial^2)^{-1} [(\partial_\sigma(k_\rho-\overline{j}_\rho) - \partial_\rho(k_\sigma - \overline{j}_\sigma)](x) \tag{7}
\end{equation}

\footnote{The solution of the tensorial PDE's linear system

a)
\begin{align*} &\varepsilon^{\alpha\beta\gamma\sigma}\,\partial_\alpha\,W_{\beta\gamma} = j_\sigma\\
&\partial_\mu\,W_{\rho\mu} = \partial_\mu\,W_{\mu\nu} = 0
\end{align*}
is formally given by
$W_{\mu\nu}=\frac{1}{12}\,(\partial^2)^{-1}(\partial_\alpha\,\widetilde{j}^{\alpha\mu\nu})$ with $\widetilde{j}^{\alpha\mu\nu} = \varepsilon_{\alpha\mu\nu\rho}\,j^\rho$.

Note that $\varepsilon^{\alpha\beta\gamma\sigma}\,\partial_\alpha (\partial_\beta a_\gamma - \partial_\gamma a_\beta) \equiv 0$.
\newline b) The path integral eq(10) can be tought as the path integral solution of eq(9) in presence of a Dirac algebra white-noise stirring ([4]).}

The whole set of global monopole generalized Maxwell equations is equivalent to the solution of the matrix (bivector) bosonic ``wave function''
\begin{equation}
[F](x) = \frac 14\,\big\{(F_{\mu\nu}- \varepsilon_{\mu\nu \rho\tau} B^{\rho\tau})[\gamma^\mu,\gamma^\nu]_-\big\} \tag{8}
\end{equation}
which by its turn is postulated to satisfies the ``Dirac'' like matricial wave equation (plus appropriate Sommerfeld radiation condidion)
\begin{equation}
(\gamma^\mu\partial_\mu)([F])(x) = j_\nu(x)\gamma^\nu - i\,\gamma_5 \gamma^\alpha k_\alpha(x) \tag{9}
\end{equation}

At this point we introduce our proposal for an (euclidean) quantum field theory for quantum electrodynamics with magnetic monopoles.

We consider as our basic field variable to be second quantized not the usual Electromagnetic potential, but the scalar-complex matricial field $[F](x)$ as given by eq(8).

Unfortunatelly, a canonical Lagrangean and a corresponding variational principle is mising for eq(9) as far as this author knows ([1]). However we overcome such difficulty by proposition the simples bosonic action leading to eq(9) at its functional extremum point.

It reads as of as
\begin{align*}
&\quad S([F],[F]^+) = \int d^4x\\
&{\rm Tr}_{\rm Dirac}\bigg\{ ((\gamma^\mu\partial_\mu)[F] - \gamma^\nu\,j_\nu - i\gamma^\mu\,\gamma_5\,k_\mu)^+
((\gamma^\mu\partial_\mu)[F] - \gamma^\nu\,j_\nu - i\gamma^\mu\,\gamma_5\,k_\mu)\bigg\}(x) \tag{10}
\end{align*}

It is obvious that the (unique!) minimum of the quadratic functional eq(10) is achieved on the classical motion eq(9), thus on the classical Maxwell equations eq(5), eq(2) with magnetic monopole source.

The second quantization of the Monopole Maxwell Equations in the Euclidean Space-Time through path integrals is now straightforward proposed as
\begin{align*}
&Z[J_{\mu\nu}(x)] = \frac{1}{Z(0)} \times \int D^F([F])D^F([F]^+) \times
\exp\bigg(-\frac 12\, S([F],[F^+])\bigg)\\
& \times\exp\bigg\{i \int d^\mu x \bigg[J_{\mu\nu}(x)\bigg(\overbrace
{\varepsilon^{\mu\nu\sigma\rho}\,B_{\sigma\rho}
+\frac 14\, {\rm Tr}_{\rm Dirac} (\gamma^\mu[F]\gamma^\nu}^{F_{\mu\nu}(x)=}\bigg)(x)\bigg]\bigg\}, \tag{11}
\end{align*}
where $D^F([F])$, $D^F([F]^+)$ denotes the Feynman product measures. Explicitly:
\begin{equation}
D^F([F](x)) = \left(\prod_{A,B=1}^4 \left(\prod_{x\in\mathbb{R}^4} d[F]_{AB}(x)\right)\right) \tag{12}
\end{equation}

\vglue .2in

\begin{equation}
D^F([F]^+(x)) = \left(\prod_{A,B=1}^4 \left(\prod_{x\in R^4} d[F]^+_{AB}(x)\right)\right) \tag{13}
\end{equation}
 Let us remark that one can ``bosonize'' eq(11) through the classical variable change ([3]) eq(8) into the path integral eq(11), under the hypothesis of its validity at the quantum level.
\begin{equation}
D^F([F](x)) = \det{}^{\frac 12} \big(\mathcal{L}_{\alpha\zeta}^{\nu\rho}\big)\times D^F[F_{\mu\nu}(x)] \tag{14}
\end{equation}
which means that now the new fundamental variable are the gauge invariant 
magnetic monopole electromagnetic field strenght $\{F_{\mu\nu}(x)\}$ instead of the original matrix field $[F](x)$.

Here, the ``tensorial matrix'' in eq(14) is  explicitly given by (it is    monopole electromagnetic field strenght independent.)
\begin{equation}
\mathcal{L}_{\alpha\zeta}^{\nu\rho} = |\gamma^\nu\gamma^\rho >\,<\gamma^\alpha\gamma^\zeta| \tag{15}
\end{equation}
\begin{equation}
\det{}^{\frac 12} \big(\mathcal{L}_{\alpha\zeta}^{\nu\rho}\big) = {\rm Tr}(\gamma^\rho\gamma^\nu\, \gamma^\alpha \gamma^\zeta) \tag{16}
\end{equation}

The free action eq(10) is now given by the Cramer-Julia action for antisymmetric rank two tensors (for vanishing sources $k^\gamma(x) = j^\gamma(x) \equiv 0$)
\begin{equation}
S^{(0)}[F_{\mu\nu}] = \frac 16 \int_{R^4} d^4x (\partial_\mu F_{\nu\rho} + \partial_\nu F_{\mu\rho} + \partial_\rho F_{\mu\nu})^2(x) \tag{17}
\end{equation}

Our path integral expression for the generating functional of the second quantized generalized magnetic monopole electromagnetic field now reads as of as:
\begin{align*}
Z[J_{\mu\nu}(x)] &= \frac{1}{Z(0)} \int D^F(F_{\mu\nu}(x)) \exp\bigg\{ - S^{(0)} [F_{\mu\nu}]\bigg\}\\
&\quad \exp\bigg\{-S^{(1)} [F_{\mu\nu}, k^\gamma,j^\gamma]\bigg\} \exp\bigg\{i \int d^4x \bigg[J_{\mu\nu}.F^{\mu\nu}(x)\bigg]\bigg\} \tag{18}
\end{align*}

\footnote{
\begin{description}
\item[1)]
\begin{equation}
\gamma^\mu\partial_\mu([F]) = (\partial_\mu F_{\mu\rho})\gamma^\rho - \frac i2\,\varepsilon^{\sigma\mu\nu\rho} \gamma_\sigma\gamma_5\partial_\mu\,F_{\nu\rho} \tag{1}
\end{equation}
\item[2)]
\begin{align*}
&{\rm Tr}_{\rm Dirac}\bigg\{\bigg[(\partial_\mu F_{\mu\rho})\gamma^\rho - \frac i2\, \varepsilon^{\sigma\mu\nu\zeta} \gamma_\sigma \gamma_5\partial_\mu F_{\nu\zeta}\bigg]^2\bigg\}
= 4(\partial^\mu F_{\mu\rho})^2 + 12[(\partial_\mu F_{\nu\rho})^2 + 
(\partial_\rho F_{\nu\rho})(\partial_\mu F_{\mu\nu})\\
&\qquad\qquad\qquad + (\partial_\nu F_{\nu\rho})(\partial_\mu F_{\rho\mu)}] \sim\, (\partial_\mu F_{\nu\rho} + \partial_\nu F_{\mu\rho} + \partial_\rho F_{\mu\nu})^2 \tag{2}
\end{align*}
\item[3)] Namely, the action eq(10) is quadratic (Gaussian) in terms of the Monopole Electromagnetic strenght fields $F_{\mu\nu}$.
\end{description}}

Here the functional weight $S^{(1)}[F_{\rho\nu},k^\gamma, j^\gamma]$ depends explicitly on the currents-sourcer generating the second quantized electromagnetic field supporting now magnetic monopolic charges. It contains terms of the form
\begin{align}
&\text{a)}\qquad\qquad \int d^4x(j^\mu\,j_\mu)(x)\qquad\qquad  \tag{19-a}\\
&\text{b)}\qquad\qquad \int d^4x(k^\mu\,k_\mu)(x)\qquad\qquad  \tag{19-b}\\
&\text{c)}\qquad\qquad \int d^4xF_{\alpha\beta}(x) \varepsilon^{\alpha\beta\sigma\zeta}(\partial_\sigma k_\zeta-\partial_\zeta k_\sigma)(x)\qquad\qquad  \tag{19-c}\\
&\text{d)}\qquad\qquad \int d^4x\, \varepsilon^{\sigma\mu\nu\zeta} (\partial_\mu\,F_{\nu\zeta})(x) k_\sigma(x)\qquad\qquad \tag{19-d}
\end{align}
, etc...

In the case for quantum sources, the second quantized Dirac electron field $\{\psi_e(x), \overline{\psi}_e(x)\}$ and the postulated second quantized monopole field $\{\Omega_M(x), \overline{\Omega}_M(x)\}$, we should introduce on eq(18) the further euclidean path integral on the matter fields and the currents (second reference - [3])
\begin{equation}
j^\mu(x) = e(\overline{\psi}_e\,\gamma^\mu\,\psi_e)(x) \tag{20}
\end{equation}
\begin{equation}
k^\mu(x) = g(\overline{\Omega}_\mu\,\gamma^\mu\,\Omega_\mu)(x) \tag{21}
\end{equation}

Note that the introduction of matter-sources for the generalized fields introduces apparently non-renormalizable thirring like interaction on the pure matter sector (see eq(19-a)-eq(19-b)). This means that Q.E.D in presence of magnetic monopoles is non-renormalizable, not satisfies the quantum mechanical unitary condition, besides of confining electrical charge$^{(*)}$.

\footnote{$^{(*)}$ On refs ([4]) we have proved that electromagnetic path integrals, when formally written in terms of vector potentials are of fourth-order, thus leading to confinement of electrical charge. This result constraint us to claim that only a new electrodynamics written directly in terms of the Electromagnetic Field strenght has chances to be consistent at a second quantized level when in presence of magnetic monopoles.} 

\bigskip

\noindent\textbf{Appendix 1} -- A new ``fermionization'' for Maxwell Equations.

Let $[\widehat{F}]$ be the complex matrix taking values on the Dirac algebra of matrixes
\begin{equation}
[\widehat{F}] = \frac 14\, F_{\mu\nu}(x)[\gamma^\mu,\gamma^\nu]_- + (\gamma^\mu)a_\mu(x) \tag{22}
\end{equation}
where $a_\mu(x)$ is an fixed external divergenceless field and the antisymmetric rank-two tensor $F_{\mu\nu}(x)$ satisfies the Maxwell equations for monopoles
\begin{equation}
\partial_\mu\,F^{\mu\nu}(x) = j^\nu(x) \tag{23}
\end{equation}
\begin{equation}
\partial_\mu(^*F^{\mu\nu})(x) = k^\nu(x) \tag{24}
\end{equation}

Then $[\widehat{F}]$ satisfies ours generalized Haddamard-Hestenes equation
\begin{equation}
\gamma^\mu\partial_\mu([\widehat{F}](x)) = \gamma^\rho f_\rho(x) + (\gamma_\sigma \gamma_5 k^\sigma(x)) + \frac 14\, [\gamma^\alpha, \gamma^\beta]_-\, F_{\alpha\beta}(a^\gamma(x)) \tag{25}
\end{equation}

Conversely, if we define the rank-two antisymmetric tensor field through the relationship below
\begin{equation}
F_{\mu\nu}(x) = \frac 14\,{\rm Tr}_{\rm Dirac}(\gamma^\mu[F]\gamma^\nu)  + (\varepsilon^{\mu\nu\sigma\rho}\,B_{\sigma\rho}(x)) \tag{26}
\end{equation}
with
\begin{equation}
\varepsilon^{\mu\nu\sigma\rho}(\partial_\mu\,B_{\sigma\rho})(x) = 0 \tag{27}
\end{equation}
\begin{equation}
(\partial_\mu\,B^{\mu\nu})(x) = k^\nu(x) - \mathcal{F}^\nu(x) \tag{28}
\end{equation}
where
\begin{equation}
\mathcal{F}^\nu(x) = \frac 18\,\varepsilon^{\mu\nu\sigma\rho}\,\partial_\mu 
\left\{ {\rm Tr}_{\rm Dirac} (\gamma^\sigma[F]\gamma^\rho)\right\} \tag{29}
\end{equation}
then
\begin{align*}
&\partial_\mu\,F^{\mu\nu}(x) = j^\nu(x)\\
&\partial_\mu(^*F^{\mu\nu}(x)) = k^\nu(x). \tag{30}
\end{align*}

\medskip

\noindent\textbf{Acknowledgments:} The author is very thankfull to CNPq for financial support for a scientific visit to IMEEC-UNICAMP and too the Mathematical Physics group of Professor W. Rodrigues.

\vglue .3in

\noindent\textbf{References}
\begin{itemize}
\item[{[1]}] - Carlo Cafaro - Finite-Range Electromagnetic Interaction and Magnetic Charges: Space-Time algebra or algebra of Physical Space? - apXiv.math-ph/0702044v2 - 15 May 2007.
\item[{-}] Daniel Zwanziger - Phys. Rev. 03 - Issue 4, pp. 880, (1971).
\item[{-}] W.A. Rodrigues Jr and E. Capelas de Oliveira, 
´´The Many Faces of Maxwell, Dirac and Einstein Equations", A Clifford Bundle Approach, Lecture Notes in Physics 722, Springer, Heidelbergerg, 2007.
\item[{[2]}] Hestenes, D. - Space-Time Algebra, New York, Gordan \& Breach - 1966.
\item[{[3]}] Luiz C.L. Botelho - Random Operators and Stochastic Equations - V. 24, pp 79-92, 2016.
\item[{-}] Luiz C.L. Botelho - Int. Journal of Theor. Physics, V. 48, pp 1554-1558, 2009.
\item[{[4]}] Luiz C.L. Botelho - International Journal of Modern Physics A, V. 15, no. 5, 755-770; 2000
\item[{-}] Luiz C.L. Botelho - IJTP, v. 48, 1701-1711; 2009.
\item[{-}] Luiz C.L. Botelho - Physical Review 70D\, 045010-1-045010-7; 2004.
\end{itemize}

\end{document}